\documentclass{aastex}

\slugcomment{}

\shorttitle{Carbon star CIT 6}
\shortauthors{Dinh-V-Trung \& Lim}

\begin{document}

\title{Tracing the asymmetry in the envelope around the carbon star CIT 6}

\author{Dinh-V-Trung\altaffilmark{1} and Jeremy Lim}
\altaffiltext{1}{On leave from Center for Quantum Electronics, Institute of Physics\\ 
Vietnamese Academy of Science and Technology, 10 DaoTan, BaDinh, Hanoi, Vietnam}
\email{trung@asiaa.sinica.edu.tw, jlim@asiaa.sinica.edu.tw}
\affil{Institute of Astronomy and Astrophysics, Academia Sinica\\ 
P.O Box 23-141, Taipei 10617, Taiwan.}

\begin{abstract}
We present high angular resolution observations of HC$_3$N J=5--4 line and 7 mm continumm emission from
the extreme carbon star CIT 6. We find that the 7 mm continuum emission is unresolved and has a flux consistent 
with black-body thermal radiation from the central star. The HC$_3$N J=5--4 line emission originates 
from an asymmetric and clumpy expanding envelope comprising
two separate shells of HC$_3$N J=5--4 emission: (i) a faint outer shell that is nearly spherical which has
a radius of 8\arcsec; and (ii) a thick and incomplete
inner shell that resembles a one-arm spiral starting at or close to the central star
and extending out to a radius of about 5\arcsec. Our observations therefore suggest that the mass
loss from CIT 6 is strongly modulated with time and highly anisotropic. Furthermore, a
comparison between the data and our excitation modelling 
results suggests an unusually high abundance of HC$_3$N in its envelope. We discuss the possibility that
the envelope might be shaped by the presence of a previously suggested possible binary companion.
The abundance of HC$_3$N may be enhanced in spiral shocks produced by the interaction between the circumstellar
envelope of CIT 6 and its companion star.
\end{abstract}

\section{Introduction}
The circumstellar envelope around carbon-rich asymptotic giant branch (AGB) stars is well known to be the site
of very active photochemistry and hence a rich source of
molecular emission lines. The prime example is the massive carbon rich circumstellar 
envelope of the nearby AGB star IRC+10216 (CW Leo), where more than 50 molecular species 
have been detected. Many spectral
line surveys in the millimeter wavelength range have been conducted toward IRC+10216 (He et al. 2008,
Cernicharo \& Gu\'{e}lin 2000, Kawaguchi et al. 1995). These surveys show that the emission lines from
cyanopolyyne molecules such as HC$_3$N and HC$_5$N are very prominent in the millimeter wavelengths. 
Interferometric observations of HC$_3$N lines in the 3-mm band by Bieging \& Tafalla 1994 with BIMA and by
Gu\'{e}lin et al. (2000) with the IRAM PdBI show that
the cyanopolyyne molecules have a hollow shell distribution, consistent with the predictions of chemical
models for a carbon rich circumstellar envelope (Cherchneff et al. 1993, Millar \& Herbt 1994, Millar et al. 2000, 
Brown \& Millar 2003). Even higher angular resolution observations of HC$_3$N J=5--4
and HC$_5$N J=16--15 lines in the 7 mm band by  Dinh-V-Trung \& Lim (2008) with the Very Large Array (VLA),
which are detectable over a more radially extended region, revealed 
the presence of a number of incomplete shells representing different 
episodes of mass loss enhancement from the central AGB star.
These observations show that the emission of cyanopolyyne molecules can be used to trace 
the small scale structures of the envelope.
 
The spatial distribution of cyanoplyyne molecules in the circumstellar 
envelope around other carbon-rich stars has been much less studied at high angular resolution.  
Here, we present observations of the HC$_3$N J=5--4 line around 
CIT 6 (RW LMi, GL 1403 or IRAS 10131+3049) with the VLA. CIT 6, a semiregular variable, is
and extreme carbon star enshrouded in a massive molecular envelope, very similar to 
the archetypical carbon star IRC+10216. It has
a pulsation period of 640 days, which, based on the period-luminosity relation for evolved
stars, translates to a distance of $\sim$400 pc (Cohen 1980 and Cohen \& Hitchon 1996). 
For easy comparison with previous work, we will adopt a distance
of 440 pc for CIT 6, which is similar to that used by Schoeier et al. (2002).
High resolution optical imaging of Schmidth et al. (2002) shows a series of arcs in
the optical reflection nebula around CIT 6, within 2\arcsec\, of the star. These arcs comprise the 
density enhancement in the envelope and indicate variations 
of mass loss on time scale of several hundred years.
Such dusty arcs have previously been seen in a number of other circumstellar envelopes
and AGB and post-AGB stars such as IRC+10216 (Mauron \& Huggins 1999), the Egg nebula (Sahai et al. 1998)
and some proto-planetary nebulae (Hrivnak et al. 2001).
 
Interferometric observations of CO J=1--0 line by Neri et al. (1998) 
and Meixner et al. (1998) at angular resolution of 3\arcsec--6\arcsec\, reveal a bright central 
core surrounded by a more diffuse but roughly spherical envelope,
expanding at a velocity $\sim$18 kms$^{-1}$. Using sophisticated 
radiative transfer models, Schoeier \& Olofsson (2001) and Schoeier et al. (2002) estimated a mass loss rate 
of 5 10$^{-6}$ M$_\odot$ yr$^{-1}$ for CIT 6 by fitting the the strength of CO rotational lines. 
They also noted that a more satisfactory match between model predictions and observations could be obtained
if mass loss variation is introduced into the model. Teyssier et al. (2006) also reached similar conclusion
from modelling the single dish observations of CO rotational lines in CIT 6. Apart from CO, several other molecules 
have been detected in CIT 6, including
cyanopolyynes (Zhang et al. 2008). The overall chemical makeup of its envelope is similar
to that of the archetypical carbon rich envelope IRC+10216, with the exception of enhanced SiC$_2$ and cyanopolyyne
emissions. The distributions of other molecules such as the cyanopolyynes
(HC$_3$N and HC$_5$N) and HCN and CN have been mapped at an
angular resolution of about 3\arcsec\, by Lindqvist et al. (2000). 
Only the emission of cyanogen radical CN was well resolved into
an incomplete and elongated expanding shell. 
 
In this paper we present our high angular resolution observations of HC$_3$N J=5--4 emission from CIT 6 
obtained with the Very Large Array (VLA\footnote{The VLA is a facility of the
National Radio Astronomy Observatory, which is operated by Associated Universities, Inc., under
contract with the National Science Foundation}). The observations allow us to probe the structure of the
envelope at arcsec scale or $\sim$6 10$^{15}$ cm.
\section{Observation}
We observe CIT 6 on 2003 April 13, using the Very Large
Array (VLA) in its most compact configuration. The telescope was pointed at 
$\alpha_{j2000}$=10h16m02.27s, $\delta_{j2000}$=30d34m18.6s, the location of the star
as listed in the SIMBAD database (Loup et al. 1993). 
The rest frequency of the HC$_3$N J=5$-$4 line 
as compiled by the Lovas/NIST database (Lovas 2004) is 45.490316 GHz. To observe this line
we configured the VLA correlator in the 2AC mode with 6.25 MHz bandwidth over
64 channels, thus providing a velocity resolution of 0.65 kms$^{-1}$ per channel over a 
useful velocity coverage of $\sim$ 40 kms$^{-1}$. The total on-source time is about 1 hour.
We monitored the nearby quasar 0958+324 at frequent intervals to correct for
the antennas gain variations caused primarily by atmospheric fluctuations. The stronger quasars 0927+390 and 1229+020 were 
used to correct for the shape of the bandpass 
and its variation with time. The absolute flux scale of our observations was determined from observation of 
standard quasar 3C 147 (0542+498).
 
We edited and calibrated the raw visibilities using the AIPS data reduction package. The calibrated visibilities
were then Fourier transformed to form the DIRTY images. We employed the robust   
weighting to obtain a satisfactory compromise between angular resolution and sensitivity. The DIRTY images were deconvolved
using normal clean algorithm implemented in AIPS, providing a synthesized beam of 1.5"x1.4" at position angle PA=24.8$^\circ$. 
The rms noise level in our channel maps of HC$_3$N J=5$-$4 emission is 6.7 mJy beam$^{-1}$ in each 
velocity channel of 2.6 kms$^{-1}$. The corresponding conversion factor between the brightness 
temperature of the HC$_3$N J=5--4 emission and 
the flux density is $\sim$ 4mJy/K.

We also configured the VLA in normal continuum mode at 7 mm. The on source integration time for CIT 6 is about 10 minutes.
The same calirators as above were used for complex gain and absolute flux density calibration. We processed the 
continuum data in the same manner as the line data. The resulting synthesized beam is 2".3x1".6 at position angle PA=75$^\circ$.
The rms noise level of the continuum map is $\sim$0.4 mJy beam$^{-1}$.
\section{Results}
We detected an unresolved continuum source at 7 mm with a flux density of 2.4 $\pm$ 0.4 mJy at
10h16m02.27s and 30d34m19.1s. This is coincident within the errors (about 0.4\arcsec) with the position of an unresolved
continuum source at 3 mm with a flux of 8 $\pm$ 0.4 mJy detected by Lindqvist et al. (2000). 
The spectral index between 7 mm and 3 mm is very close to 2 (i.e. S$_\nu$ $\sim\,\nu^2$),
consistent with stellar black body radiation from presumably the central AGB star in CIT 6.
At shorter wavelengths, the spectral index is larger with S$_\nu$ $\sim\,\nu^{3.1}$, indicating significant contribution from
dust in the envelope around CIT 6 (Neri et al. 1998, Marshall et al. 1992).

Figure 1 shows the channel maps of the HC$_3$N J=5--4 emission. Figure 2 shows the HC$_3$N J=5--4 profile
derived by integrating over a region where the emission is detected above 2$\sigma$ level. 
The HC$_3$N J=5--4 line has been previously observed by Fakasaku et al. (1994)
with the Nobeyama 45m telescope, which has a primary beam of about 40\arcsec\, at FWHM. 
Using the main beam efficiency provided by Fakasaku et al. (1995), we estimate a conversion factor of 4 Jy/K, thus
giving a peak flux density of about 1.7 Jy for the HC$_3$N J=5--4 line. 
This is comparable to the peak flux density of $\sim$1.6 Jy measured in our observation (see Figure 2).
Furthermore, the shape of the line profiles in both observations are also very similar, suggesting that
our VLA observation has recovered most of the emission in the HC$_3$N J=5--4 line present in the
abovementioned single dish observation. 

The channel maps of HC$_3$N J=5--4 emission shows the usual pattern of an expanding envelope, with the emitting region
being largest at the systemic velocity V$_{\rm LRS}$ = $-$2 kms$^{-1}$ and becoming progressively 
more compact at large blueshifted and redshifted velocities. From the velocity range covered by the 
emission in the channel maps we estimate that the expansion velocity of the HC$_3$N shell 
is about $\sim$17 kms$^{-1}$, which is in good agreement to the expansion velocity of 
18 kms$^{-1}$ inferred from previous CO observations (Neri et al. 1998, Meixner et al. 1998).
Unlike the well-defined hollow shell structure seen in HC$_3$N J=5--4 and HC$_5$N J=16--15 for IRC+10216
(Dinh-V-Trung \& Lim 2008), however, the spatial distribution of HC$_3$N J=5--4 emission
in CIT 6 is much more complex. In the velocity channel at V$_{\rm LSR}$=13.5 kms$^{-1}$, the emission appears 
roughly spherically symmetric. In the
velocity channels between $-$17.4 to 8.3 kms$^{-1}$, however, the emission resembles 
an incomplete shell with a mostly complete eastern portion whereas
the western part is missing. Previous observations of HC$_3$N J=10--9 line at lower angular resolution
of $\sim$3\arcsec\, by Lindqvist et al. (2000) also show a lopsided envelope that is strongly enhanced in the eastern
portion together with significant emission at the stellar position.

To show more clearly the structures within the expanding envelope of CIT 6, we integrate the line intensity spanning
velocities $-$7.1 to 5.7 kms$^{-1}$, namely the 6 velocity channels straddling the systemic velocity. 
The resulting integrated intensity map is shown in Figure 3.
The envelope clearly doest not resemble a spherical expanding hollow shell expected for molecules such as HC$_3$N. 
Instead, starting from the stellar position, the brightest portion of the emission extends to the south, curls to 
the east and north, and then curls again to the west and south, thus creating a structure resembling a one-arm spiral. 
In the channel maps, this spiral structure can be most
clearly seen at the systemic velocity of $-$2 kms$^{-1}$ and the adjacent channel at 0.6 kms$^{-1}$. 
Such a spiral distribution in
molecular emission has never been seen before in any circumstellar envelope. 

Beyond the outermost radius of the one-armed spiral, there is fainter emission tracing a nearly spherical shell with
radius of $\sim$8\arcsec. This shell can be seen, along with the one-armed spiral in the channel maps
between velocities of 3.2 and 5.7 kms$^{-1}$. The faint outer shell is more visible in the 
integrated intensity map shown in Figure 3. The outer shell appears to be centered on the 
central AGB star as indicated by its continuum emission at 7 mm and seems to be even more clumpy
than the one-armed spiral. For clarity, we sketch the location and the outline of these two structures in 
Figure 4.

Using the conversion factor of 4 mJy/K we estimate that the brightness temperature of the HC$_3$N J=5--4 emission in 
the channel maps around the systemic velocity (--2 and 0.6 kms$^{-1}$) ranges from about 5 K to 12 K. 
We note that the measured brightness temperature 
is smaller but quite significant in comparison to the kinetic temperature of of the molecular gas in the range 20 to 80 K expected from 
the modelling of the envelope of CIT 6 (Schoeier et al. 2002, Dinh-V-Trung et al. 2008 in preparation), indicating
that the optical depth in the HC$_3$N J=5--4 line is significant if the line is close to thermalization.
\section{Discussion}
\subsection{Comparison with previous observations}
The existence of spatially distinct and incomplete molecular shells as traced by the HC$_3$N J=5--4 emission
again emphasizes the similarity between CIT 6 and the archetypical carbon star IRC+10216. In the later
case, numerous incomplete expanding shells have been detected and linked
to different episodes of mass loss enhancement from the central AGB star (Dinh-V-Trung \& Lim 2008). 
In CIT 6 the presence of the shells as well as the dust arcs seen in the optical (Schmidt et al. 2002), all point to previous 
episodes of mass loss enhancement. From the radius of
the outer shell of 8 arcsec and an expansion velocity of 18 km s$^{-1}$, we estimate that the episode of mass loss
that gave rise to this thin shell 
happened about 10$^3$ years ago. The inner shell, which resembles a one-armed spiral, spans radii of $\leq$1\arcsec\, 
to 5\arcsec. 
That corresponds to a timescale between $\leq$120 to 600 years.

High resolution optical images of Schmidt et al. (2002) show a bipolar morphology for the nebula around
the envelope. Because the bipolar morphology is typical of post-AGB objects such as the Egg nebula or 
CRL 618, they suggested that CIT 6 has reached the end of the AGB phase. Our data, however, do not
show any evidence of bipolar morphology in the molecular envelope. The
bipolar morphology seen in the optical is therefore likely an illumination effect due to the peculiar distribution of 
the material close to star. For example, a spherical envelope with its polar 
caps removed will allow starlight to escape more easily in the polar directions thus creating
an apparent bipolar morphology in the reflection nebula.

\subsection{Excitation of HC$_3$N}
The existence of HC$_3$N J=5--4 emission close to the central star of CIT 6 is surprising given the commonly accepted formation
pathway for HC$_3$N and also the
excitation conditions in the envelope. Chemical models for carbon rich circumstellar 
envelope
predict that HC$_3$N molecules form from the reaction between CN, which is a photodissociation product of parent molecule
HCN, and acetylene C$_2$H$_2$:\\
CN + C$_2$H$_2$ ---$>$ HC$_3$N + H \\
Therefore, very few HC$_3$N molecules are expected to exist in the inner region of the envelope as few interstellar UV photons
can penetrate into the inner region to produce CN radicals through photodissociation of HCN. That conclusion can be
seen in the predictions of chemical models of Cherchneff et al. (1993), Millar \& Herbst (1994) and Millar et al. (2000). 
Observations of high-lying transitions of HC$_3$N observations
by Audinos et al. (1994), however, suggest that the abundance of HC$_3$N in the inner region of the envelope around the 
archetypical carbon star IRC+10216 is quite significant, nearly an order of magnitude higher than predicted by chemical models.

Do we expect to observe the HC$_3$N J=5--4 line in the inner region of the envelope ? 
The answer is likely negative because the higher gas density and kinetic temperature in the inner region can easily
lead to the excitation of HC$_3$N to higher rotational levels. That effectively reduces the optical depth, and consequently,
the intensity of low lying rotational transitions. As a result, only high lying transitions of HC$_3$N might be detected in the
inner region. Observationally, high lying transitions of HC$_3$N up to J=29--28 have been detected in the spectral line survey toward
CIT 6 of Zhang et al. (2008). Based on this simple argument, the existence of low lying HC$_3$N J=5--4 emission 
close to the central star is not expected, except when the abundance of HC$_3$N is extremely high. 

To check the above arguement more quantitatively and to explore an
alternative explanation for the observed HC$_3$N J=5--4 emission close to the AGB star of CIT 6, 
we have performed the excitation calculations for HC$_3$N molecules in the envelope of CIT 6.
The calculations were carried out using the large velocity gradient (LVG) approximation, which is justified
for the case of CIT 6 where the expansion velocity ($\sim$ 18 kms$^{-1}$) is large. We use the HC$_3$N molecular data 
compiled by Lafferty \& Lovas (1978) and Uyemura et al. (1982). We include in our model the radiative excitation of the molecule 
due to IR pumping through the $\nu_5$ bending vibrational state. As suggested by Bieging \& Tafalla (1993), this
mode is strongest and its wavelenth is close to the peak of the continuum spectrum of IRC+10216 and also
CIT 6. Therefore the $\nu_5$ bending mode should play a dominant role in the radiative excitation process. 
We include all rotational levels in the ground state and $\nu_5$=1
vibrational state up to J=30. Because the
Q-branch ro-vibrational transitions do not effectively contribute to the radiative excitation of the HC$_3$N molecule, 
we consider only transitions in the P and R-branches in our model.
We also adopt the same dipole moment of 0.18 Debye for the vibrational transition between $\nu_5$ state and the ground state 
as used in Audinos et al. (1994). The collisional cross sections between HC$_3$N and H$_2$ are assumed to follow 
the prescription of
Deguchi et al. (1984). From detailed modelling of the structure of the envelope (Dinh-V-Trung et al. 2008
in preparation, Schoeier et al. 2002), the mass loss rate of CIT 6 is found to be 5 10$^{-6}$ M$_\odot$ yr$^{-1}$. 
The gas temperature as a function of radial distance in the envelope as derived from our modelling results can be
approximated as T$_K$(r) = 33 K (r/3 10$^{16}$ cm)$^{-0.6}$. We note that the kinetic temperature in the
envelope of CIT 6 is significantly lower in comparison to that in IRC+10216 (Bieging \& Taffala 1993, Audinos et al. 1994). 
Because both CIT 6 and IRC+10216 share very similar properites such as the total luminosity, the overall shape of 
the SED (Lindqvist et al. 2000), we follow Bieging \& Tafalla (1993) and adopt a blackbody IR continuum source having 
5.1 10$^{14}$ cm in radius and a temperature of 600 K.

We adopt an abundance of HC$_3$N with respect to H$_2$ of the form f$_{HC_3N}$ = f$_0\,$exp[--4ln2(r -- r$_0$)$^2$/FWHM$^2$],
where f$_0$ is the peak abundance, $r_0$ is the radius of the peak abundance and FWHM is the full 
width at half maximum of the radial distribution.
This functional form of the abundance distribution is similar to that used by Bieging \& Tafalla (1993) to model
the HC$_3$N J=10--9 line in the envelope of IRC+10216. 
Because the HC$_3$N J=5--4 emission in CIT 6 exists over a large range of radii, 
between $\leq$1\arcsec\, to 8\arcsec\,, we use a representative value $r_0$=4 10$^{16}$ cm or 6\arcsec\, in angular
distance. 

In order to reproduced the measured brightness temperature of HC$_3$N J=5--4 line in CIT 6, which is in the
range of 5 - 12 K for channel maps around the systemic velocity, we find that a peak 
abundace of f$_0$ = 5 10$^{-6}$ is needed. In Figure 5 we show the results of our model. 
For the case with FWHM $\sim$2 10$^{16}$ cm, 
which reproduces the broad distribution of of HC$_3$N derived by Audinos et al. (1994) for IRC+10216, 
the predicted peak brightness temperature of HC$_3$N J=5--4 line is $\sim$10 K, comparable to the 
abovementioned observed brightness temperature. The predicted brightness temperature is found to peak at a smaller radius
in comparison to the underlying abundance distribution of HC$_3$N. That can be easily understood because the
location of the peak in brightness temperature is determined not just by the abundance but by the overall balance between
gas density, the abundance of HC$_3$N and the gas temperature. At the radii
between 1\arcsec\, to 2\arcsec, because of very low abundance of HC$_3$N, the predicted brightness temperature is almost zero.
To reproduce the brightness temperature of $\sim$ 5 K seen close to the central AGB star in the channel maps around
the systemic velocity, we need to broaden the abundance distribution of HC$_3$N by increasing the parameter FWHM to 3.5 10$^{16}$ cm.
The abundance of HC$_3$N in the inner envelope is then increased by more than two order of magnitude as shown in Figure 4. 
The predicted brightness temperature between the radii of 1\arcsec\, to 2\arcsec\, is now comparable to the observed value.

The qualitative calculations presented here clearly indicate that the abundance of HC$_3$N in the envelope
of CIT 6 is significantly higher in comparison to that found in the archetypical carbon rich envelope of IRC+10216, which
has a peak HC$_3$N abundance of f$_0$ = 10$^{-6}$ derived by Audinos et al. (1994).
The especially elevated abundance in the inner envelope close to the central AGB star of CIT 6 can not easily explained by the
current chemical models of carbon rich envelopes. 

\subsection{The binary hypothesis}

The inferred high abundance of HC$_3$N in the envelope around CIT 6, especially in the inner region close to the
star and the unusual spatial distribution of HC$_3$N J=5--4 emission resembling a one-arm spiral suggest that 
mechanisms other than photochemistry might be at work to form the HC$_3$N molecules. 

Gu\'{e}lin et al. (1999) noted that many molecular species in IRC+10216 are co-spatial even though they are
predicted by chemical models (Cherchneff et al. 1993, Millar et al. 2000) to form at different radial distances. 
The similar spatial distribution of different molecular species indicates that the molecules 
are all formed in a very short timescale, of the order of hundred of years. They suggest that
other mechanisms such as desoprtion from dust grain or the release from grain surface due to shocks might
be responsible for the formation of molecules in carbon rich circumstellar envelopes. 

Binary companions around AGB stars have been suggested to play important role in shaping the structure and
influencing the wind dynamics within the circumstellar envelope.
Indeed, in the hydrodynamic simulations of Mastrodemos \& Morris (1999), Edgar et al. (2008), 
the interaction with a binary companion can induce
spiral shocks and enhance the density structure in the envelope around a mass losing star. The density structure in
the envelope is predicted to resembles a one-arm spiral (see Figures 1\&2 in Edgar et al. 2008). Such spiral structure has recently
been seen in the high resolution optical image of an extreme carbon star CRL 3068 (Morris et al. 2006, Mauron \& Huggins 2006).
The elevated density and temperature expected in the spiral shocks (Edgar et al. 2008) might be conducive 
to the formation of large carbon chain molecules including HC$_3$N as suggested by Gu\'{e}lin et al. (1999).

In the case of CIT 6, optical spectropolarimetric observations of Schmidt et al. (2002) 
showed that its optical spectrum possesses a strong and featureless blue continuum exccess. 
They attributed the blue continuum exccess to the presence of a companion star of spectral type A-F buried in the envelope.
Thus the high asymmetric envelope of CIT 6 as traced by the HC$_3$N emission might be caused by the binary companion.  
Assuming that the HC$_3$N emission traces the spiral shock produced by the companion, the period of the binary system 
might be estimated from shape of the one-arm spiral. As can be
seen from the sketch of the structures in the envelope around CIT 6 (see Figure 4), the spiral makes 
almost a complete turn, starting from the central AGB star in the South-West quadrant
and ending in the North-West quadrant at the radial distance of about 5\arcsec. The dynamical timescale corresponding to
the inter-arm spacing of the spiral shock is directly related to the reflex motion of the AGB star, i.e. the orbital period
of the binary system (Mastrodemos \& Morris 1999). Using the expansion velocity of 18 kms$^{-1}$ and 
the adopted distance of 440 pc, the orbital 
period of the binary system is about 600 years. The long orbital period indicates that the companion must be
in a wide orbit around the AGB star (the orbital separation is about 70 au if the primary AGB star has a mass of 1M$_\odot$). 
The estimate of orbital period for the binary system in CIT 6 is
comparable to that (830 years) inferred by Mauron \& Huggins (2006) for the binary system in CRL 3068. Based on this
argument, we think that the hypothesis of a binary system in CIT 6 is plausible as it could naturally account for
the spatial distribution and the high abundance of HC$_3$N as seen in our observations. Future high angular resolution
observations of dense and warm gas tracers such as high J transition of CO or HCN molecules will tell whether the
spiral shock induced by the binary companion really exists within the envelope of CIT 6.

\section{Conclusion}
We have imaged at high angular resolution the distribution of HC$_3$N J=5--4 emission from 
the envelope around carbon star CIT 6. We found that the emission of HC$_3$N J=5--4 traces
(1) a faint outer spherical shell located at a radial distance of 8 arcsec and 
centered at the position of the AGB star CIT 6 revealed by the detection of 7 mm continuum emission. 
(2) a thick and incomplete inner shell resembling
a one-arm spiral. The presence of multiple shells in CIT 6 suggests that the mass loss from this
star is highly anisotropic and episodic. From excitation modelling of HC$_3$N molecules we inferred
that the abundance of HC$_3$N is unusually high in CIT 6 in comparison to the well known 
carbon star IRC+10216. We suggest that the observed spatial distribution of the emission and the
inferred high abundance of HC$_3$N might be caused by the presence of a binary companion in a wide
orbit around CIT 6. 
\acknowledgements
We thank the VLA staff for their help with the observations. This research has made use of 
NASA's Astrophysics Data System Bibliographic Services
and the SIMBAD database, operated at CDS, Strasbourg, France.

\newpage

\begin{figure*}[ht]
\plotone{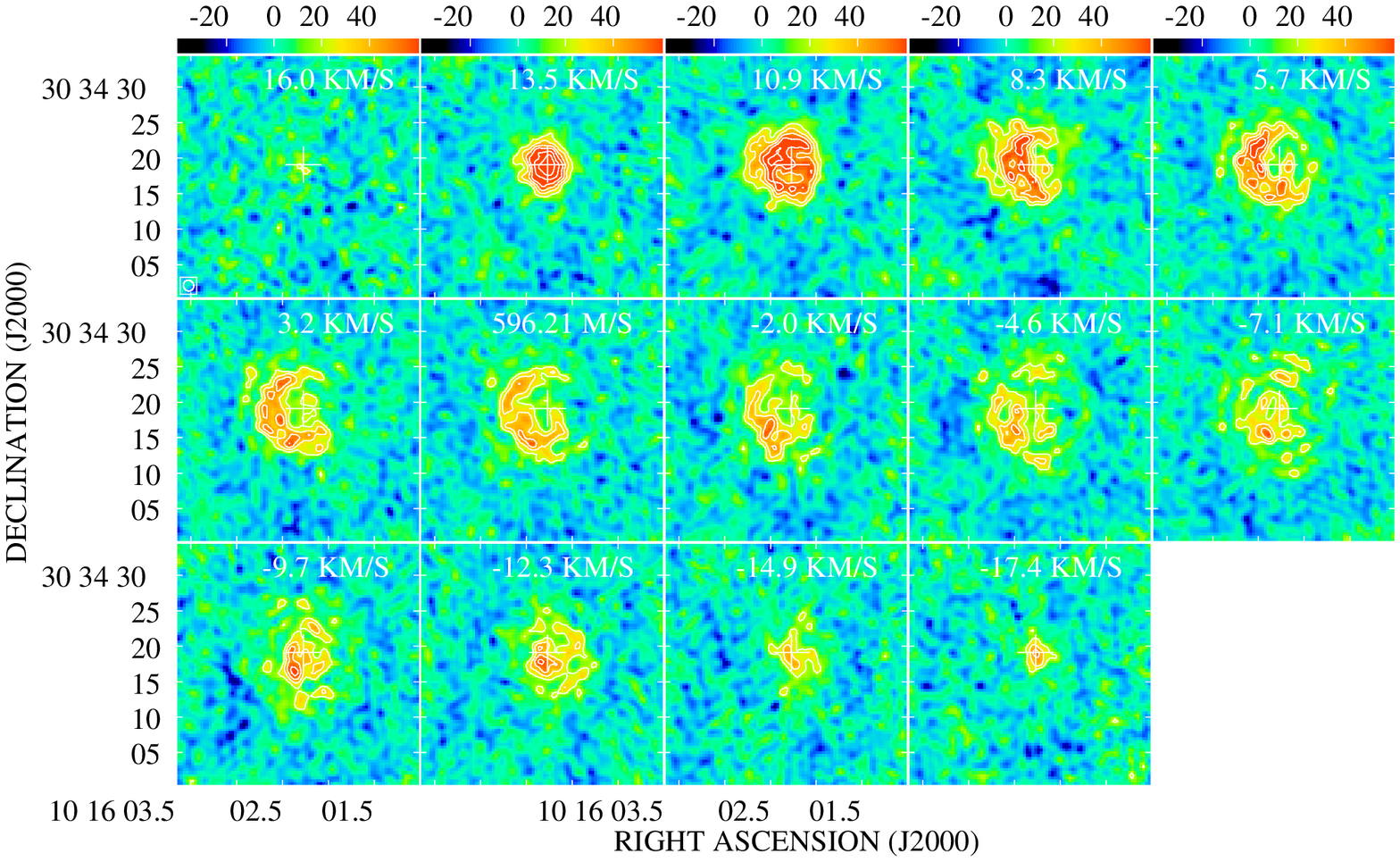}
\caption{Channel maps of the HC$_3$N J=$5-$4 emission (shown in false color and also in contours). The LSR velocity
is indicated in the upper right of each frame. The contour levels are 
(3, 5, 7, 9, 12 and 15)$\sigma$ with $\sigma$=6.7 mJy beam$^{-1}$. The synthesized beam is shown in lower left corner
of the upper left frame. The corresponding conversion factor 
between the brightness temperature and flux density is 4 mJy/K.}
\label{fig1}
\end{figure*}

\begin{figure*}[ht]
\plotone{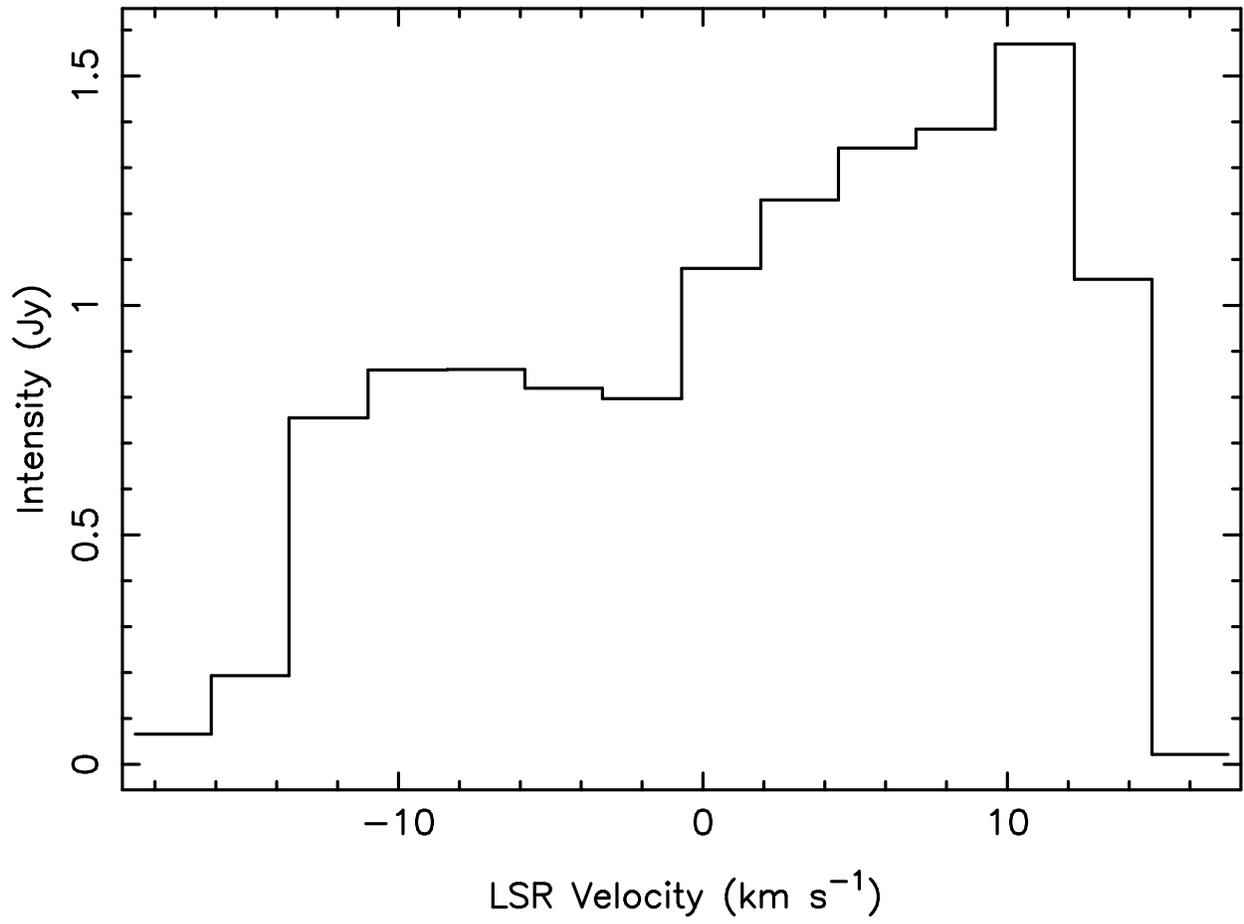}
\caption{Total intensity profile of the HC$_3$N J=$5-$4 line from CIT 6.}
\label{fig2}
\end{figure*}

\begin{figure*}[ht]
\plotone{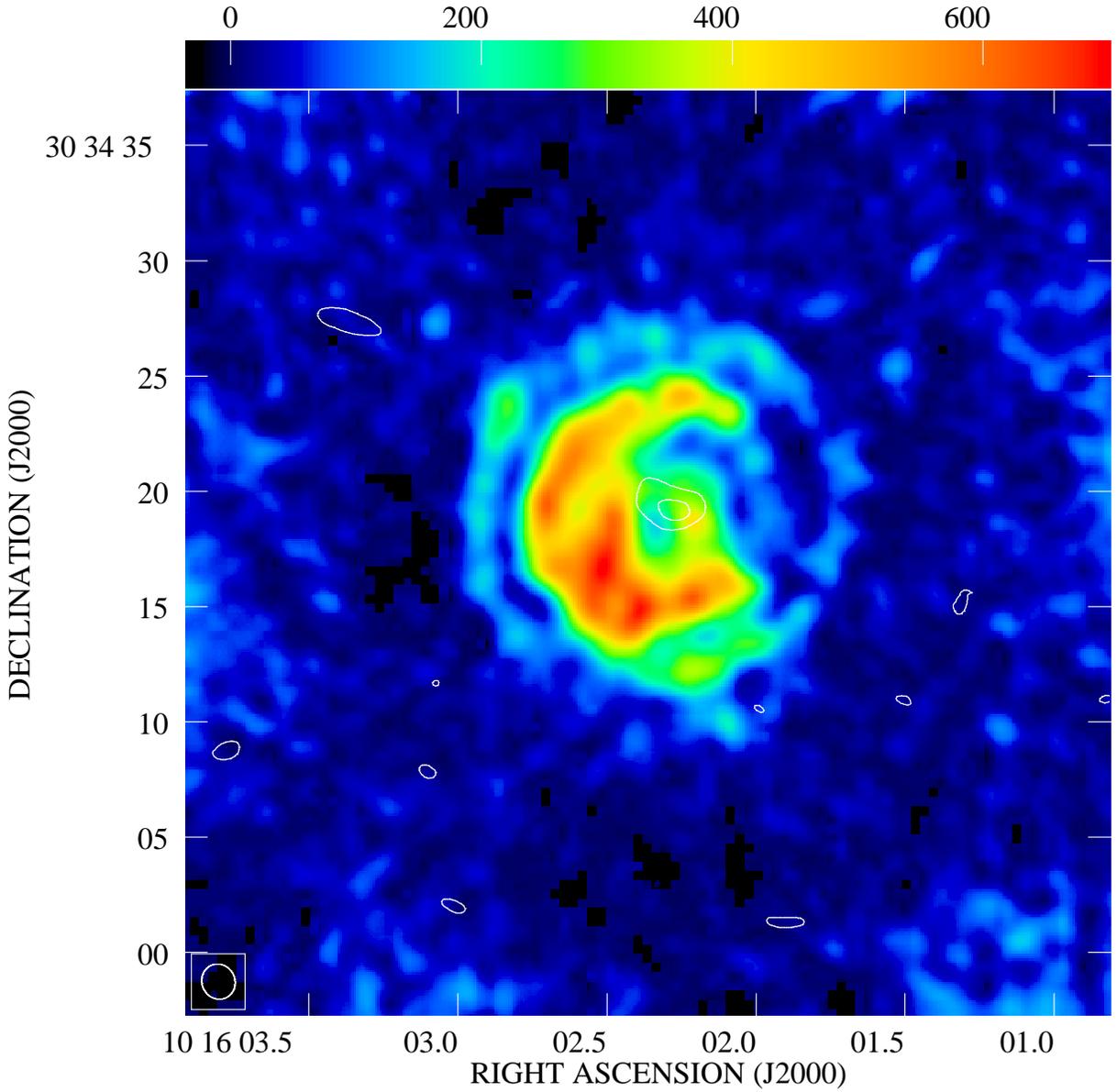}
\caption{The intensity of the HC$_3$N J=$5-$4 emission (shown in false color) integrated between velocity
channels $-$7.1 to 5.7 kms$^{-1}$. The 7 mm 
continuum emission is also shown in contours. The contour levels are 1, 2 mJy beam$^{-1}$. The intensity scale is
mJy kms$^{-1}$/beam. An outer nearly spherical
shell and an incomplete shell resembling a one-arm spiral starting from the location of the central star
indicated by the continuum emission are evident.}
\label{fig3}
\end{figure*}

\begin{figure*}[ht]
\plotone{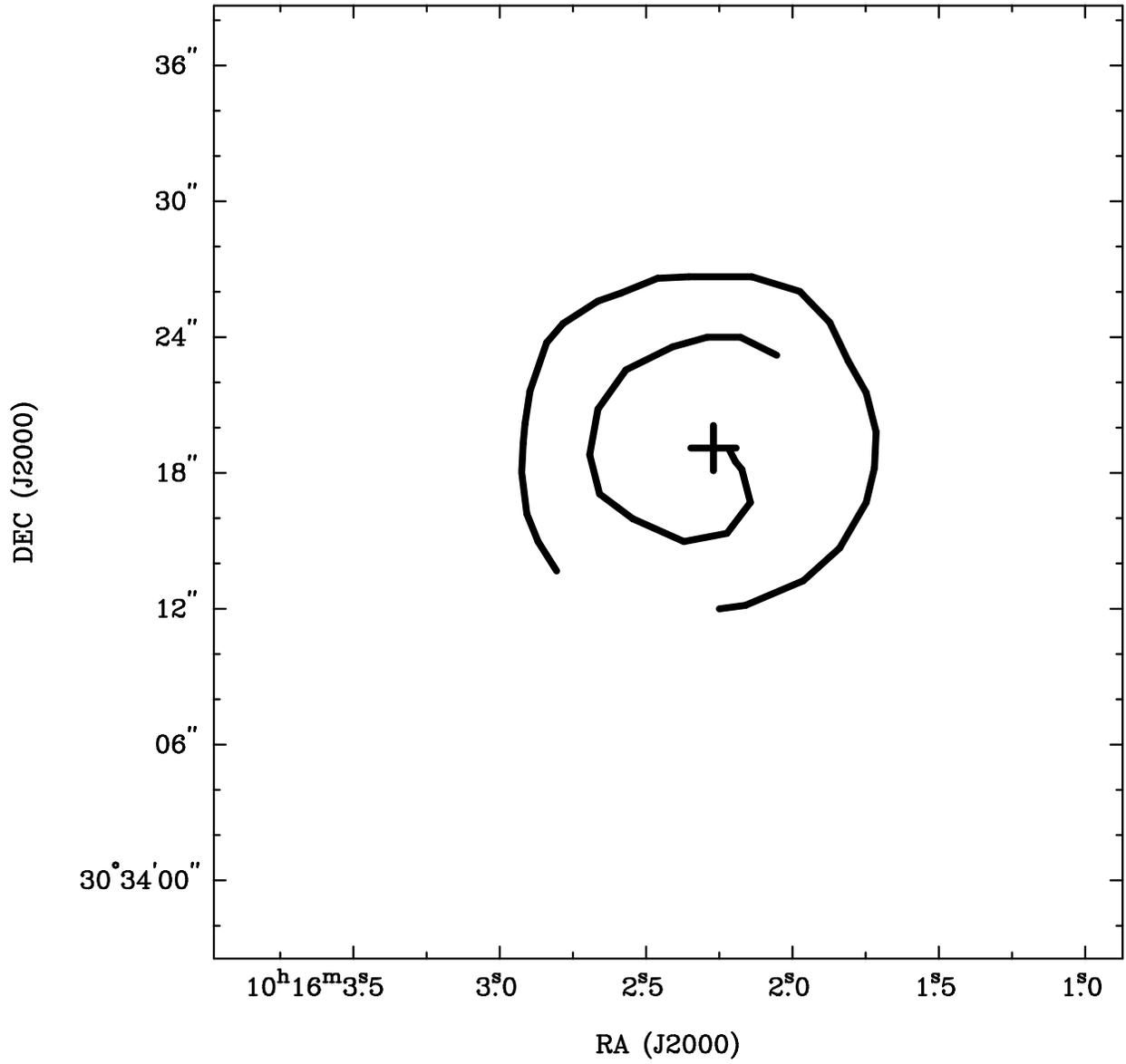}
\caption{Sketch of the outer spherical thin shell and the one-arm spiral in CIT 6. The cross denotes
the position of the AGB star.}
\label{fig4}
\end{figure*}

\begin{figure*}[ht]
\plotone{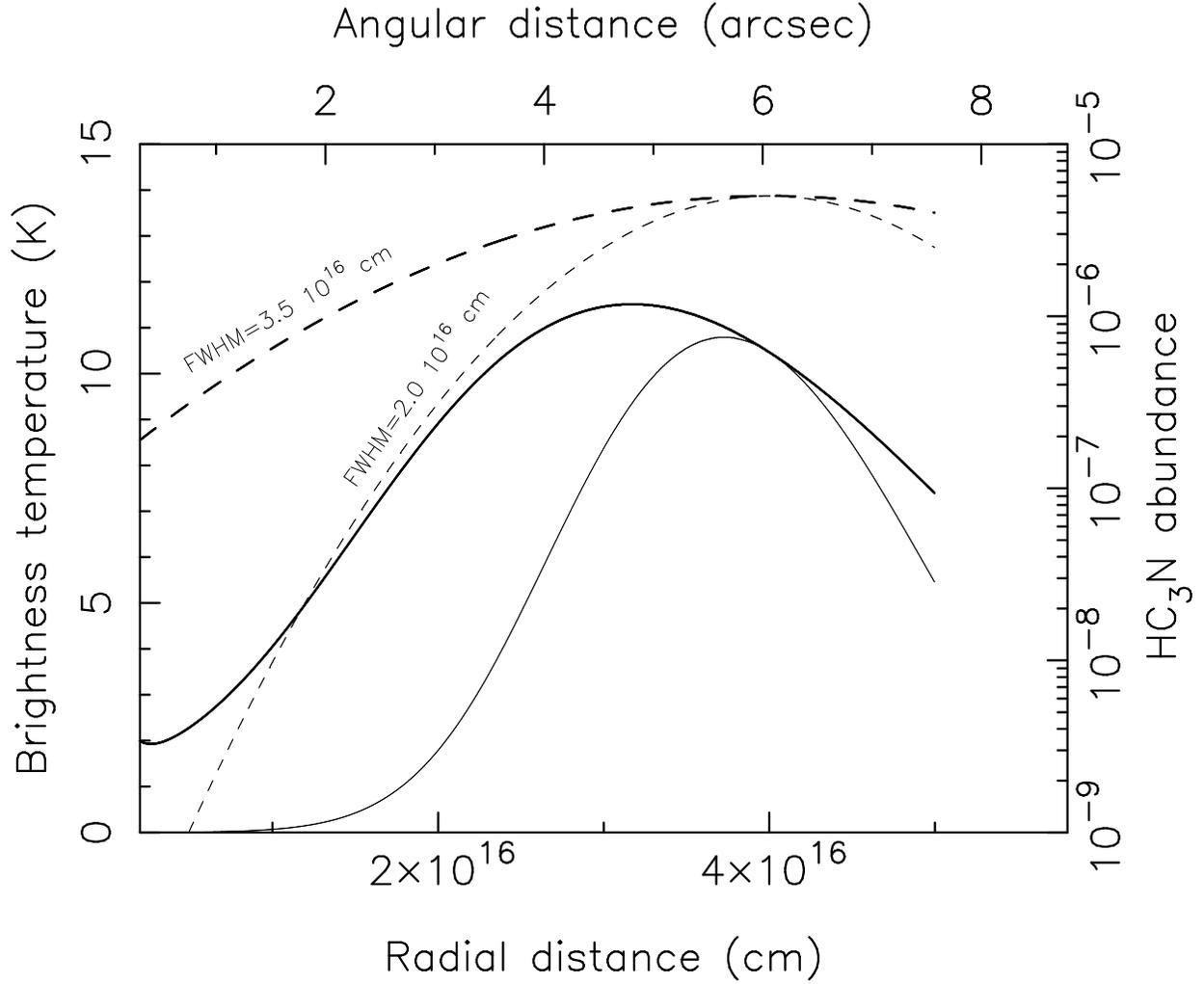}
\caption{Model calculation of the brightness temperature of HC$_3$N J=5--4. The dashed lines represent the
abundance of HC$_3$N with respect to H$_2$ in the envelope with different values of the FWHM parameter. 
The solid lines show the calculated brightness 
temperature. Thin lines represent the model with FWHM = 2 10$^{16}$ cm while the thick lines represent the model
with broader distribution FWHM = 3.5 10$^{16}$ cm, respectively.}
\label{fig5}
\end{figure*}
\end{document}